\documentclass[11pt,twoside]{article}
\usepackage{./asp2014}

\aspSuppressVolSlug
\resetcounters

\begin{document}

\title{ESA Sky: a new Astronomy Multi-Mission Interface}
\author{Bruno Mer\'in$^1$, Jes\'us Salgado$^1$, Fabrizio Giordano$^1$,
Deborah Baines$^1$, Mar\'ia-Henar Sarmiento$^1$, Bel\'en L\'opez Mart\'i$^1$,
Elena Racero$^1$, Ra\'ul Guti\'errez$^1$, Andy Pollock$^2$, Michael Rosa$^1$,
Javier Castellanos$^1$, Juan Gonz\'alez$^1$, Ignacio Le\'on$^1$, I\~naki Ortiz de Landaluce$^1$,
Pilar de Teodoro$^1$, Sara Nieto$^1$, Daniel J. Lennon$^1$, Christophe Arviset$^1$,
Guido de Marchi$^3$, William O'Mullane$^1$
\affil{$^1$European Space Astronomy Centre, Camino Bajo del Castillo s/n, Villanueva de la Ca\~nada, 28692 Madrid
Spain \email{Bruno.Merin@esa.int}}
\affil{$^2$Sheffield University, Sheffield, UK}
\affil{$^3$Science Support Office, ESTEC/SRE-SA, Keplerlaan 1, 2201 AZ Noordwijk, The Netherlands}
}

\paperauthor{Bruno~Mer\'in}{Bruno.Merin@esa.int}{orcid.org/0000-0002-8555-3012}{ESA}{SRE}{Madrid}{Madrid}{28692}{Spain}
\paperauthor{Jes\'us Salgado}{}{}{ESA}{SRE}{Madrid}{Madrid}{28692}{Spain}
\paperauthor{Fabrizio Giordano}{}{}{ESA}{SRE}{Madrid}{Madrid}{28692}{Spain}
\paperauthor{Deborah Baines}{}{}{ESA}{SRE}{Madrid}{Madrid}{28692}{Spain}
\paperauthor{Mar\'ia-Henar Sarmiento}{}{}{ESA}{SRE}{Madrid}{Madrid}{28692}{Spain}, 
\paperauthor{Bel\'en L\'opez Mart\'i}{}{}{ESA}{SRE}{Madrid}{Madrid}{28692}{Spain}
\paperauthor{Elena Racero}{}{}{ESA}{SRE}{Madrid}{Madrid}{28692}{Spain} 
\paperauthor{Ra\'ul Guti\'errez}{}{}{ESA}{SRE}{Madrid}{Madrid}{28692}{Spain} 
\paperauthor{Andy Pollock}{}{}{University of Sheffield}{}{Sheffield}{}{}{UK} 
\paperauthor{Michael Rosa}{}{}{ESA}{SRE}{Madrid}{Madrid}{28692}{Spain}
\paperauthor{Javier Castellanos}{}{}{ESA}{SRE}{Madrid}{Madrid}{28692}{Spain} 
\paperauthor{Juan Gonz\'alez}{}{}{ESA}{SRE}{Madrid}{Madrid}{28692}{Spain}
\paperauthor{Ignacio Le\'on}{}{}{ESA}{SRE}{Madrid}{Madrid}{28692}{Spain} 
\paperauthor{I\~naki Ortiz de Landaluce}{}{}{ESA}{SRE}{Madrid}{Madrid}{28692}{Spain}
\paperauthor{Pilar de Teodoro}{}{}{ESA}{SRE}{Madrid}{Madrid}{28692}{Spain} 
\paperauthor{Sara Nieto}{}{}{ESA}{SRE}{Madrid}{Madrid}{28692}{Spain} 
\paperauthor{Daniel J. Lennon}{}{}{ESA}{SRE}{Madrid}{Madrid}{28692}{Spain}
\paperauthor{Christophe Arviset}{}{}{ESA}{SRE}{Madrid}{Madrid}{28692}{Spain}
\paperauthor{Guido de Marchi}{}{orcid.org/0000-0002-8555-3012}{ESA}{SRE}{Noordwijk}{}{2201 AZ}{The Netherlands} 
\paperauthor{William O'Mullane}{}{}{ESA}{SRE}{Madrid}{Madrid}{28692}{Spain}

\begin{abstract}
We present a science-driven discovery portal for all the ESA Astronomy Missions called ESA Sky that allow users to explore the multi-wavelength sky and to seamlessly retrieve science-ready data in all ESA Astronomy mission archives from a web application without prior-knowledge of any of the missions. The first public beta\footnote{\url{http://archives.esac.esa.int/esasky-beta/}} of the service has been released, currently featuring an interface for exploration of the multi-wavelength sky and for single and/or multiple target searches of science-ready imaging data and catalogues. Future releases will enable retrieval of spectra and will have special time-domain exploration features.\\
From a technical point of view, the system offers progressive multi-resolution all-sky projections of full mission datasets using a new generation of HEALPix projections called HiPS, developed at the CDS; detailed geometrical footprints to connect the all-sky mosaics to individual observations; and direct access to science-ready data at the underlying mission-specific science archives.
\end{abstract}

\section{Introduction and motivation}

With the goal of maximizing the exploitation of science data obtained with telescopes and infrastructures made possible with public funding, organizations are striving to make their data holdings more accesible to scientists around the world.

In order to define the best possible interface to the data from ESA Astronomy missions, in 2014, ESA polled scientists at its ESAC and ESTEC Establishments for the ingredients of an ideal interface to data. With the responses it became clear that users needed to have a simple and fast access to selected high-quality data for all missions without having to know all technical details of those missions.

\section{Application design}

To fulfill the requests from the users, a web application was designed, that sits in front of all the mission archives and presents and serves to the user selected data from those mission archives in a simple way (see Fig. \ref{fig:ESASky_Design}). This architecture allows three main use cases: i) exploration of the multi-wavelength skies, ii) data retrieval for single targets and iii) data retrieval for target lists.

\articlefigure{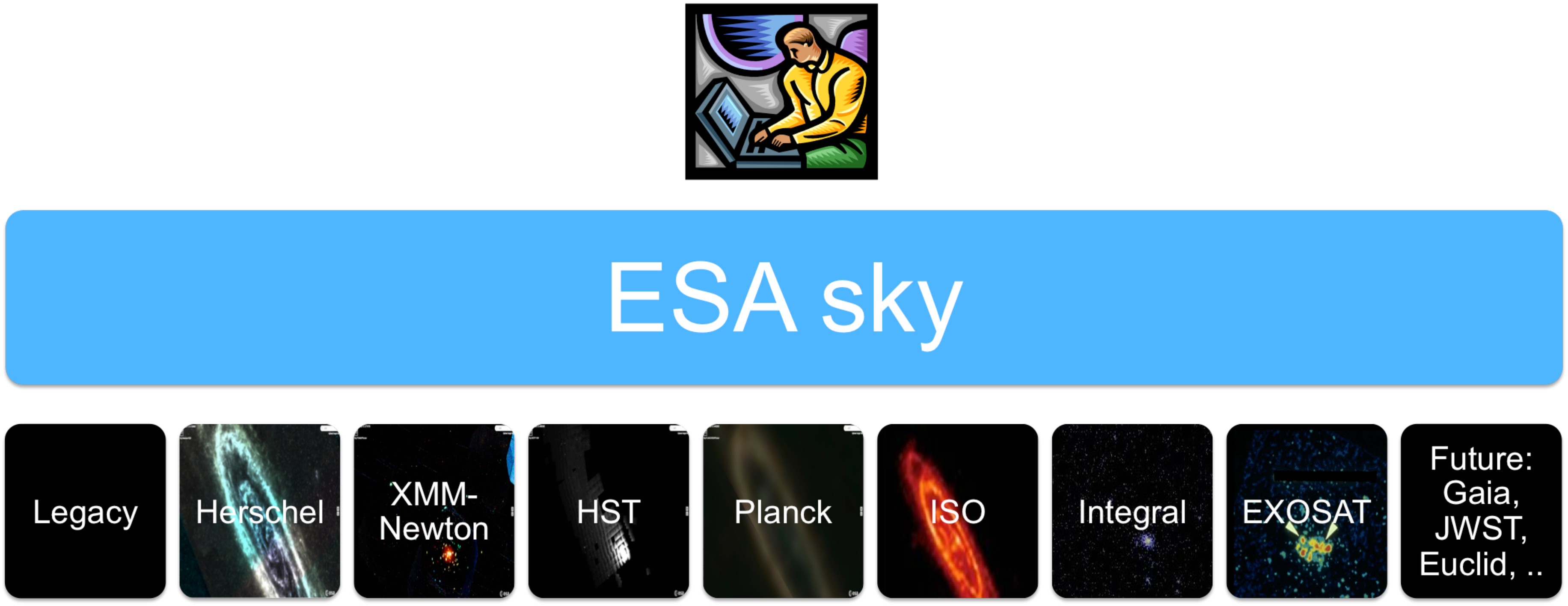}{fig:ESASky_Design}{Design of the ESA Sky application. A web application sits in between the user and the mission archives, selecting files from the archives and serving them to the user in a simplified way.}

From a technical point of view, it the system offers progressive multi-resolution all-sky projections of full mission datasets using a new generation of HEALPix projections called HiPS \citep{Fernique2015}, developed at the CDS; detailed geometrical footprints to connect the all-sky mosaics to individual observations; and direct access to science-ready data at the underlying mission-specific science archives.

The user is shown a sky in optical and allowed to explore by zooming and panning to any place of the sky, choose which wavelength or frequency to look at (from $\gamma$-ray to radio wavelengths) and search for single targets or target lists.

\section{Beta release}

With the occasion of the ADASS XXV conference in Sydney, we have made public the first beta release of the application to gather feedback from professional astronomers and astronomical data centers before the first official public release, scheduled a few months after the beta release. The address of the beta version of the application is: \\

\url{http://archives.esac.esa.int/esasky-beta}\\

\articlefigure{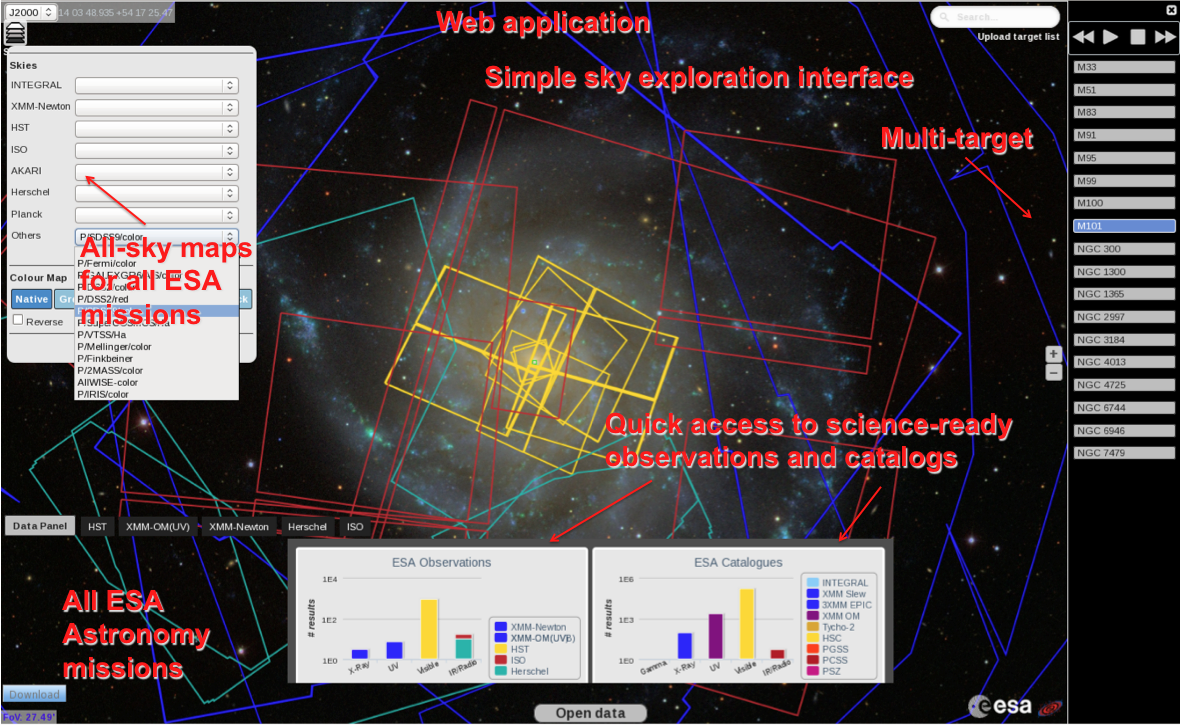}{fig:ESASky_Features}{Features of ESA Sky. The top left panel lets the user choose any sky to visualize from $\gamma$-ray to radio wavelengths, the top right allows the user to search for any object by name, coordinates or upload a target list. The panel at the bottom displays the results of the automatic queries to the mission archives for observations and/or catalogues in the area visualized.}

\subsection{Data holdings}

This beta version contains the imaging datasets and catalogues from the following ESA Astronomy missions :

\begin{itemize}
\item 13 years of data in the Gamma-ray domain from INTEGRAL\footnote{\url{http://www.esa.int/esaSC/120374_index_0_m.html}}: all-sky maps and 2077 $\gamma$-ray sources.
\item 16 years of data in the X-ray, UV and optical domain from XMM-Newton \footnote{\url{http://sci.esa.int/xmm-newton/}}: all sky maps, 8721 observations, over 560,000 X-ray and 6 million optical/UV sources.
\item 25 years of data in the UV, optical and Near-IR from HST\footnote{\url{https://en.wikipedia.org/wiki/Hubble_Space_Telescope}}: all-sky maps, 588820 observations and $\sim$ 29 million UV/optical/NIR sources.
\item 4 years of data in the optical from Hipparcos\footnote{\url{https://www.rssd.esa.int/SA/HIPPARCOS/docs/vol2_all.pdf}}: a catalog with $\sim$ 2.5 million optical sources.
\item 2.3 years of data in the infrared from the Infrared Space Observatory ISO \citep{Kessler1996}: one all sky map and 47652 observations.
\item 4 years of infrared data from the Herschel Space Observatory \citep{Pilbratt2010}: 8 all sky maps and 16039 observations.
\item 4 years of infrared and radio data from Planck \citep{Tauber2010}: 9 all-sky maps and 153142 infrared and/or radio detections.
\end{itemize}

In particular for this release, ESA has produced HiPS all sky maps of all the datasets in the list above with the expert support from the mission specialists at ESAC, in some cases also combining several filters into RGB color-composite maps. These maps represent a new and unique way of visualizing the whole history of observations in just one single multi-resolution hierarchical map.

\subsection{Documentation, feedback and contributions}

Even if we have designed it with simplicity in mind, interested users can also find a full set of documentation and video tutorials at: \\

\url{http://www.cosmos.esa.int/web/esdc/esasky-help}\\

\noindent We encourage feedback from users through the helpdesk of the ESAC Science Data Center:\\

\url{https://support.cosmos.esa.int/esdc/}\\

\noindent Astronomy data centers wishing to link their data holdings through ESA Sky are welcome to do so and should consult the simple instructions at: \\

\url{http://www.cosmos.esa.int/web/esdc/esasky-contributing}

\section{Future plans}

ESA's future plans for the application include giving access to spectroscopic data and adding special features to explore the time-domain component of the data holdings. 


\acknowledgements  We acknowledge the excellent support from Pierre Fernique and Thomas Boch from the Centre de Donn\'ees Astronomiques de Strasbourg (France) and from the expert science and technical staff at ESAC for the creation of this service, in particular from Pedro Rogr\'iguez (XMM-Newton Science Operations Centre), Guillaume Belanger (INTEGRAL Science Operations Centre), Alejandro Lorca (Computer Support Group) and Roberto Prieto (Computer Support Group). This work also benefited from experience gained from projects supported by the SRE-OO research funding.

\bibliography{biblio}

\begin{thebibliography}{}
\expandafter\ifx\csname natexlab\endcsname\relax\def\natexlab#1{#1}\fi
\expandafter\ifx\csname url\endcsname\relax
  \def\url#1{\texttt{#1}}\fi
\expandafter\ifx\csname urlprefix\endcsname\relax\def\urlprefix{URL }\fi
\providecommand{\eprint}[2][]{\url{#2}}

\bibitem[{{Fernique} et~al.(2015){Fernique}, {Allen}, {Boch}, {Oberto},
  {Pineau}, {Durand}, {Bot}, {Cambr{\'e}sy}, {Derriere}, {Genova}, \&
  {Bonnarel}}]{Fernique2015}
{Fernique}, P., {Allen}, M.~G., {Boch}, T., {Oberto}, A., {Pineau}, F.-X.,
  {Durand}, D., {Bot}, C., {Cambr{\'e}sy}, L., {Derriere}, S., {Genova}, F., \&
  {Bonnarel}, F. 2015, \aap, 578, A114. \eprint{1505.02291}

\bibitem[{{Kessler} et~al.(1996){Kessler}, {Steinz}, {Anderegg}, {Clavel},
  {Drechsel}, {Estaria}, {Faelker}, {Riedinger}, {Robson}, {Taylor}, \&
  {Xim{\'e}nez de Ferr{\'a}n}}]{Kessler1996}
{Kessler}, M.~F., {Steinz}, J.~A., {Anderegg}, M.~E., {Clavel}, J., {Drechsel},
  G., {Estaria}, P., {Faelker}, J., {Riedinger}, J.~R., {Robson}, A., {Taylor},
  B.~G., \& {Xim{\'e}nez de Ferr{\'a}n}, S. 1996, \aap, 315, L27

\bibitem[{{Pilbratt} et~al.(2010){Pilbratt}, {Riedinger}, {Passvogel}, {Crone},
  {Doyle}, {Gageur}, {Heras}, {Jewell}, {Metcalfe}, {Ott}, \&
  {Schmidt}}]{Pilbratt2010}
{Pilbratt}, G.~L., {Riedinger}, J.~R., {Passvogel}, T., {Crone}, G., {Doyle},
  D., {Gageur}, U., {Heras}, A.~M., {Jewell}, C., {Metcalfe}, L., {Ott}, S., \&
  {Schmidt}, M. 2010, \aap, 518, L1. \eprint{1005.5331}

\bibitem[{{Tauber} et~al.(2010){Tauber}, {Mandolesi}, {Puget}, {Banos},
  {Bersanelli}, {Bouchet}, {Butler}, {Charra}, {Crone}, {Dodsworth}, \&
  et~al.}]{Tauber2010}
{Tauber}, J.~A., {Mandolesi}, N., {Puget}, J.-L., {Banos}, T., {Bersanelli},
  M., {Bouchet}, F.~R., {Butler}, R.~C., {Charra}, J., {Crone}, G.,
  {Dodsworth}, J., \& et~al. 2010, \aap, 520, A1

\end{thebibliography}

%
%

\end{document}